\documentclass[letterpaper,10pt]{article} 
\usepackage{osameet3} %% use version 3 for proper copyright statement

\usepackage{amsmath,amssymb}
\usepackage{physics,siunitx}
\usepackage[colorlinks=true,bookmarks=false,citecolor=blue,urlcolor=blue]{hyperref} %pdflatex
\usepackage[labelfont=bf]{caption}
\usepackage{ragged2e}
\usepackage{epsfig}

\begin{document}
\title{Wavelength-accurate and wafer-scale process for nonlinear frequency mixers in thin-film lithium niobate}

\author{
    C.~J.~Xin$^{1,\dagger,*}$,
    Shengyuan~Lu$^{1,\dagger}$,
    Jiayu~Yang$^{1,\dagger}$,
    Amirhassan~Shams-Ansari$^{1,2,\dagger}$,
    Boris~Desiatov$^{1,3}$,
    Letícia~S.~Magalhães$^{1}$,
    Soumya~S.~Ghosh$^{1,4}$,
    Erin McGee$^{1,5}$,
    Dylan~Renaud$^{1}$,
    Nicholas~Achuthan$^{1}$,
    Arseniy~Zvyagintsev$^{1}$,
    David~Barton~III$^{1,6}$,
    Neil~Sinclair$^{1}$,
    Marko~Lončar$^{1,*}$,
}

\address{
    $^1$ John A. Paulson School of Engineering and Applied Sciences, Harvard University, Cambridge, MA, USA \\
    $^2$ DRS Daylight Solutions, 164565 Via Esprillo, San Diego, CA, USA \\
    $^3$ Faculty of Engineering, Bar-Ilan University, Ramat-Gan, Israel \\
    $^4$ Department of Physics, Harvard University, Cambridge, MA, USA \\
    $^5$ Department of Physics, Stevens Institute of Technology, Hoboken, NJ, USA \\
    $^6$ Department of Materials Science and Engineering, Northwestern University, Evanston, IL, USA \\
    $^\dagger$ These authors contributed equally \\
    $^*$ Corresponding authors: 
        {\rm \href{mailto:loncar@seas.harvard.edu}{loncar@seas.harvard.edu}}; 
        {\rm \href{mailto:cxin@g.harvard.edu}{cxin@g.harvard.edu}}
}
%%Uncomment the following line to override copyright year from the default current year.
%\copyrightyear{2024}

\begin{abstract*} 
    Recent advancements in thin-film lithium niobate (TFLN) photonics have led to a new generation of high-performance electro-optic devices, including modulators, frequency combs, and microwave-to-optical transducers. 
    However, the broader adoption of TFLN-based devices that rely on all-optical nonlinearities have been limited by the sensitivity of quasi-phase matching (QPM), realized via ferroelectric poling, to fabrication tolerances. 
    Here, we propose a scalable fabrication process aimed at improving the wavelength-accuracy of optical frequency mixers in TFLN. 
    In contrast to the conventional pole-before-etch approach, we first define the waveguide in TFLN and then perform ferroelectric poling. 
    This sequence allows for precise metrology before and after waveguide definition to fully capture the geometry imperfections. 
    Systematic errors can also be calibrated by measuring a subset of devices to fine-tune the QPM design for remaining devices on the wafer. 
    Using this method, we fabricated a large number of second harmonic generation devices aimed at generating $737~\si{\nm}$ light, with $73\%$ operating within $5~\si{\nm}$ of the target wavelength. 
    Furthermore, we also demonstrate thermo-optic tuning and trimming of the devices via cladding deposition, with the former bringing $\sim96\%$ of tested devices to the target wavelength. 
    Our technique enables the rapid growth of integrated quantum frequency converters, photon pair sources, and optical parametric amplifiers, thus facilitating the integration of TFLN-based nonlinear frequency mixers into more complex and functional photonic systems.
\end{abstract*}

\section*{Introduction} 
The thin-film lithium niobate (TFLN) on insulator integrated photonics platform has experienced rapid growth within the last decade due to the commercial availability of high-quality ion-sliced substrates, as well as the development of optimized etching processes for fabricating low-loss, nanophotonic waveguides \cite{Zhang2017-dm}.
Underpinning these developments are the excellent material properties of lithium niobate (LN) including a wide transparency window, spanning near-ultraviolet to mid-infrared, and strong second-order nonlinearities ($\chi^{(2)}$).\autocite{Weis1985-qw}
This has enabled many technologically relevant devices, such as electro-optical modulators and optical frequency mixers that convert light between widely separated wavelength bands. 
Compared to their bulk counterparts \autocite{Abouellell1989-hj, Bazzan2015-ar, Myers1995-uv}, TFLN devices feature significant improvements in key figures of merit such as modulation speed  \autocite{Xu2022-ib, Kharel2021-tg}, optical bandwidth \autocite{Javid2021-ce}, and efficiency \autocite{Lu2020-nv, Chen2024-yt}, due to high optical mode confinement in nanophotonic waveguides.
Moreover, reduced device footprints in TFLN and wafer-scale processing \autocite{Luke2020-gp} using standard microfabrication techniques provide a solid foundation for mass production and integration of many individual components into more complex photonic circuits.
For instance, circuits comprising arrays of electro-optic modulators in TFLN have already been used to demonstrate functionalities such as large-scale matrix-vector multiplications \autocite{Lin2023-vv}.

On the other hand, $\chi^{(2)}$-based optical frequency mixers, which are used in applications ranging from frequency up-/down-conversion \autocite{Boes2023-kv} to optical parametric amplification \autocite{Jankowski2022-qn,Guo2022-lh} to quantum light sources \autocite{Saravi2021-at}, have thus far been implemented mostly as standalone components or in relatively simple circuits featuring at most two mixers operating in concert \autocite{Park2024-zh,Hwang2023-ra,Sjaardema2022-ha}.
This is despite ongoing interest in developing these components for applications such as quantum frequency conversion for interfacing with solid state single photon emitters and memories \autocite{Riedel2023-ks} and multiplexed photon pair sources \autocite{Morrison2022-kt}.  

One obstacle towards the scaling and integration of $\chi^{(2)}$-based optical frequency mixers in TFLN is the extreme sensitivity of conventional quasi-phase matched (QPM) three-wave mixing processes—namely, sum- and difference-frequency generation (SFG and DFG), and spontaneous parametric downconversion (SPDC)—to device geometry \autocite{Fejer1992-gc}.
QPM is implemented by periodically inverting domains of TFLN waveguide which changes the sign of $\chi^{(2)}$.
The momentum of such grating is then used to compensate for mismatch between wave vectors of different waves participating in the nonlinear mixing process.
In particular, nanometer-level local geometry variations in the waveguide cross-section can lead to a reduction in device efficiency at even moderate device lengths \autocite{Kuo2022-fm,Zhao2023-fq}.
Recent work has shown that this issue can be effectively mitigated by using extensive metrology of the initial thickness of the device layer to perform fine adjustments to the QPM grating design \autocite{Chen2024-yt}.
Additionally, a less metrology-intensive mitigation approach based on post-fabrication trimming using spatially varying thermo-optic tuning has also been demonstrated, albeit at the cost of high power consumption and sacrificing low-temperature compatibility \autocite{Li2023-ls}.

However, in larger samples (e.g. wafers), long range waveguide geometry variations become a salient issue in addition to the aforementioned local geometry fluctuations. 
This drift arises from imperfections in the dry etch process---often using an argon-based physical etch---that is used to realize TFLN devices, and can cause shifts in the operating wavelength of devices well beyond the thermo-optic tuning range.
Thus, for applications requiring accuracy in the operating wavelength—such as for interfacing with atom-like single-photon emitters that have fixed emission wavelengths (e.g. silicon-vacancy center in diamond at $737~\si{\nm}$ with $\sim 80~\si{\GHz}$ demonstrated tuning range \autocite{Machielse2019-na})—the conventional approach is to sweep the design parameters to guarantee at least one device that meets specifications \autocite{Jankowski2021-hf,Hwang2023-kp}.
While this is viable for standalone devices and small photonic circuits, the number of trial circuits that must be fabricated to guarantee one meets specifications scales geometrically with the number of frequency mixing devices in each circuit even if one assumes unity yield on all other components in the circuit.

Here, we propose a scalable process for $\chi^{(2)}$-based optical frequency mixers in $x$-cut TFLN featuring high wavelength accuracy.
In our approach, QPM gratings are defined after waveguide etching, allowing for fine-tuning of the grating parameters by taking into account variations in waveguide geometry parameters.
Furthermore, systematic errors can be calibrated out using a small subset of calibration devices defined on the same wafer.
The approach is experimentally validated by realizing second harmonic generation (SHG) mixers producing 737 nm light. This wavelength corresponds to the optical transition of silicon-vacancy color center in diamond, a leading solid-state quantum memory \autocite{Bhaskar2020-db}.
We show that $73\%$ of devices out of $84$ tested across a $4''$ wafer support SHG of $(737 \pm 5)~\si{\nm}$. Furthermore, we demonstrate on a separate wafer that wavelength trimming and tuning, using SiO$_{2}$ deposition and thermal tuning respectively, can bring $96\%$ of tested devices within the same range.   

\section*{Conventional pole-before-etch-process for quasi-phase matching in TFLN}
\begin{figure}[h]
    \caption{\justifying \bf
        Etch-before-pole fabrication process flow for nonlinear frequency mixers in thin-film lithium niobate (TFLN).
    } \label{main:fig1}
    \includegraphics[width=\textwidth]{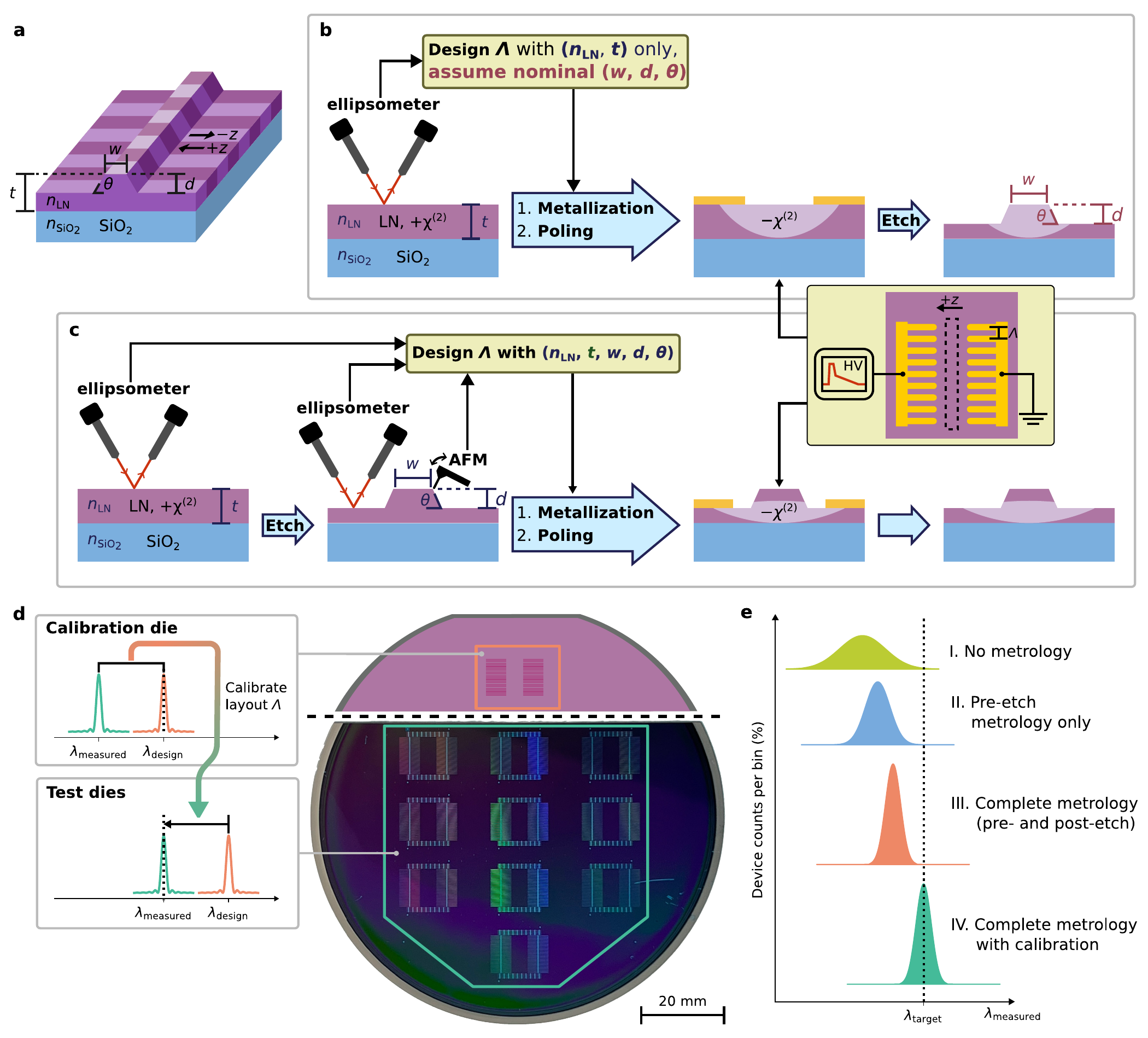}
    \caption*{\justifying
        \textbf{a},~Illustration of a conventional quasi-phase matched (QPM) waveguide for parametric three-wave mixing in $x$-cut TFLN. 
        Alternating dark and light bands along the waveguide correspond to the alternating signs of optical $\chi^{(2)}$. 
        The QPM period, $\Lambda$,  required to phase match three wavelengths of interest is dependent on the LN (silica) material index, $n_{\rm LN (SiO_2)}$, as well as the waveguide cross-section. 
        The latter crucially depends on TFLN thickness, $t$; etch depth, $d$; top width, $w$; and sidewall angle, $\theta$. 
        We note that QPM designs are far less sensitive to $n_{SiO_2}$ than the remaining labeled parameters; nonetheless, we label it here for completeness. 
        \textbf{b},~Schematic of conventional pole-before-etch fabrication process flow, where $n_\text{LN}$ and t are the only measured parameters. 
        Ferroelectric poling via a high voltage (HV) pulse applied to coplanar finger electrodes is used to pattern the QPM grating before etching to define the waveguide profile (inset). 
        \textbf{c},~Schematic of etch-before-pole process proposed here. Film thickness is measured both pre- and post-waveguide etching.
        Atomic force microscopy (AFM) is used to evaluate the waveguide cross-section profile.
        $\Lambda$ thus can be designed using complete information of the waveguide cross-section geometry. 
        Ferroelectric poling is carried out by applying a HV pulse to coplanar electrodes, with the resulting domain inversion only occurring in the waveguide slab (inset).
        \textbf{d},~Schematic of systematic error calibration carried out using a small subset of total devices (left) with an example of how this is implemented on a $4''$ wafer (right). 
        Dotted lines indicate the location of the target operating wavelength, $\lambda_0$. Calibration devices, comprising $<20\%$ of the total relevant QPM devices in each wafer run, are positioned near the wafer flat, separated from the remaining devices by dicing, and measured first to calibrate out systematic errors. 
        $\Lambda$ only needs to be swept in the calibration devices and can be fixed at the calibrated $\Lambda$ for the target process.
        \textbf{e},~Illustrations of the how distribution of device operating wavelengths is expected to deviate from the target wavelength, $\lambda_\text{target}$, when QPM grating periods are designed with (I)~no metrology information, (II)~only pre-etch film thickness, (III)~complete measured waveguide cross-section geometry, and (IV)~complete measured waveguide cross-section geometry with systematic error calibration.
    }
\end{figure}

The most basic $\chi^{(2)}$-based optical frequency mixers in TFLN rely on a rib waveguide oriented orthogonal to the material’s crystallographic z-axis (Fig.~\ref{main:fig1}a). 
For efficient parametric three-wave mixing to occur in the device, the vacuum wavelengths, $\lambda_i$, and effective propagation constants, $k_i = 2\pi n_{\text{eff}, i} / \lambda_i$ ($i \in \{1,2,3\}$) (where $n_{\text{eff}, i}$ is the effective index for $\lambda_i$), of the three interacting guided modes must satisfy energy conservation, $\lambda_3^{-1} = \lambda_1^{-1}  + \lambda_2^{-1}$, and momentum conservation, $\Delta k \equiv k_1 + k_2 - k_3 = 0$. 
The latter condition is also known as “phase matching” and is not, in general, satisfied for most processes due to material and modal dispersion of waveguide modes of interest. 
However, in LN---and other ferroelectric materials---a nonlinear three-wave mixing process can be “quasi-phase matched” by periodically flipping the sign of the second order coefficient, $\chi^{(2)}$, with period, $\Lambda\qty(n_\text{LN},t,w,d,\theta) = 2\pi/\qty(\Delta k(n_\text{LN},t,w,d,\theta))$.\autocite{Langrock2006-jk}  
The grating period depends on material indices of LN (buried oxide) layer at the wavelengths of interest, $n_{\rm LN(SiO_2)}$; TFLN thickness, $t$; waveguide rib height, $d$; waveguide top width, $w$; and sidewall angle, $\theta$ (Fig.~\ref{main:fig1}a). 
In $x$-cut TFLN, which is the crystal orientation of choice for compatibility with high-performance electro-optic devices, the QPM grating is typically fabricated by applying a strong electric field along the crystallographic $-z$ direction using coplanar finger electrode gratings with a period that matches the desired QPM grating (Fig.~\ref{main:fig1}b, inset); this is known as “ferroelectric poling”\autocite{Scrymgeour2009-zw}. 
This creates a QPM grating with spatial frequency, $K=2\pi/\Lambda$, along the direction of propagation to compensate for the phase mismatch.

In the majority of works featuring QPM three-wave mixing devices, poling takes place near the beginning of the fabrication process, \textit{before} the waveguide is defined, and thus without the knowledge of the values of $d$, $w$, and $\theta$.
The TFLN waveguide is defined using dry etching, which results in $\pm 10~\si{\nm}$ uncertainty for $d$ and $w$, and $\pm\ang{2}$ for $\theta$ (Fig.~\ref{main:fig1}b). 
Therefore, to guarantee devices operating at desired wavelengths, fabrication uncertainties are compensated for by sweeping the poling period across many devices to guarantee phase-matching over a wide wavelength range\autocite{Jankowski2021-hf, Hwang2023-kp}. 
This, however, results in a low yield and makes it very difficult to fabricate more complex nonlinear devices~(Fig.~\ref{main:fig1}e). 

\section*{Etch-before-pole process for wavelength-accurate frequency mixers in TFLN}
In this work, we propose an alternative fabrication process in which ferroelectric poling to define the QPM grating occurs \textit{after} waveguide etching~(Fig.~\ref{main:fig1}c). 
In this case, wafer-scale ellipsometry can be used both before and after etching to determine $t$ and $d$, respectively, as well as to obtain the material dispersion, $n_\text{LN}$. 
Furthermore, atomic force microscopy (AFM) can also be carried out to precisely measure the cross-sectional profile of the waveguide and thus obtain $\theta$ and $w$. 
Thus, the etch-before-pole process allows for QPM gratings to be designed with full knowledge of the waveguide geometry and material parameters. 
Assuming perfectly accurate metrology tools, this allows for all devices on a wafer to operate at the target wavelength. 
Thus, the proposed fabrication process can be both high-yield and robust to fabrication imperfections. 

In practice, however, systematic errors can be present in every parameter used to calculate the phase mismatch. 
Conveniently, the etch-before-pole process flow allows a small subset of calibration devices to be completed and characterized to calibrate the metrology tools (ellipsometer and AFM) themselves (Fig.~\ref{main:fig1}d). Figure~\ref{main:fig1}e provides an illustration of how different metrology and calibration steps impact the distribution of operating wavelengths of devices relative to a target operating wavelength, $\lambda_0$. 

\section*{Experimental results}
To evaluate the efficacy of our etch-before-pole approach, in enhancing device yield while reducing the necessity for sweeping QPM grating periods, we fabricated a total of $660$ ($360$) QPM waveguides targeting SHG of $737~\si{\nm}$ light on a wafer implementing periodic (aperiodic) poling. 
All the fabrication and metrology steps up to the extraction of calibration devices are performed on $4''$ 5\% MgO-doped TFLN wafers. 
Calibration devices are positioned near the wafer flat so that they can easily be separated from the wafer by dicing (Fig.~\ref{main:fig1}d). 

Ellipsometry performed before and after waveguide etching shows $\sim 20~\si{\nm}$ variations in $t$ and $d$ (corresponding to “pre-etch thickness” and “film etched” respectively in Fig.~\ref{main:fig2}a) across the wafer. 
Film thicknesses are sampled at multiple points along every waveguide (Fig.~\ref{main:fig2}b). 
A relatively fine sampling grid of $100~\si{\micro\m}$ along each waveguide was chosen to allow us to compare the anticipated yield for periodic QPM gratings and aperiodic QPM grating designs, discussed below. 
In the latter, poling periods are adapted to geometry fluctuations along a single device as in Ref.~\cite{Chen2024-yt}. 
We note that although local etch depth fluctuations along a single waveguide are only on the order of several nanometers, long-range variation across the full wafer is an order of magnitude larger. 
This indicates that while the assumption of uniform etch depth may be valid at chip scale which is that used by most demonstrations to date, it may not necessarily hold for larger sample sizes, which could be used for novel device geometries and co-integration of PPLN. 

In addition to film thickness measurements, the waveguide cross-section profile is measured via tapping mode AFM at multiple points across the wafer, with one witness structure measured per 60 QPM waveguides. 
Combining the AFM traces with film thickness maps provides complete knowledge of the waveguide geometry both locally (along each QPM waveguide) and across the entire wafer. 
This information is then used in finite difference eigenmode (FDE) simulations to find the phase mismatch $\Delta k^{\rm (sim)}$ for the SHG of $\lambda_0 = 737~\si{\nm}$ light. 
Both the fundamental and second harmonic occupy the fundamental pseudo-transverse electric (TE00) guided mode. 

Next, we theoretically investigate possible strategies to design QPM gratings and compare their ability to produce SHG close to $\lambda_0$ (Fig.~\ref{main:fig3}a). 
In the absence of measured substrate and fabrication imperfections Method I assumes the same and uniform QPM grating period (calculated for $t = 600~\si{\nm}$, $w = 1.5~\si{\micro\m}$ , $d = 300~\si{\nm}$, and $\theta = \ang{60}$) applied along each waveguide on the wafer. 
However, due to variations in LN thickness and waveguide geometry, this approach would result in waveguides with SHG spread across $\sim 60~\si{\nm}$ range around $\lambda_0$, resulting in poor yield. 
Leveraging the knowledge of measured $t$ to design QPM gratings considerably narrows the distribution of SHG produced by different waveguides. 
Two approaches are considered: periodic (Method~IIa) which uses uniform $\Lambda(t_0)$ based on TFLN thickness ($t_0$) sampled at a single point in the middle of the waveguide, and aperiodic (Method~IIb) which uses $\Lambda\qty[t(y)]$, the function of the position along the waveguide length ($y$) due to measured thickness variations obtained from measurements. 
However, in both cases, the center of the distribution is displaced from $\lambda_0$ due to the unknown $n_{\rm LN}$, $d$, $w$, and $\theta$.
Finally, when measured values are used for all waveguide parameters, the distribution of SHG generated by different waveguides can be re-centered on the $\lambda_0$. 
This is true even when a periodic QPM grating $\Lambda(n_{\rm LN}, t_0, w_0, d_0, \theta_0)$ is used (Method~IIIa) based on a single waveguide geometry sample in the middle of the waveguide.
The width of the distribution is also expected to narrow compared to Method~IIa since fluctuations from device to device, due to non-uniform etch depths, are accounted for.
Finally, when aperiodic poling is used, by taking into account all measured parameters along each waveguide (Method~IIIb) to obtain the QPM grating $\Lambda\qty[n_{\rm LN}, t(y), w(y), d(y), \theta(y)]$, all waveguides on the wafer should produce SHG at $737~\si{\nm}$. 

In summary, QPM grating design based on the conventional pole-before-etch process results in wide distributions of SHG wavelengths produced by different waveguides (I), even when the knowledge of TFLN thickness is available (IIa, IIb). 
On the other hand, the etch-before-pole process reported in this work allows complete information of the waveguide cross-section geometry to be considered for the QPM grating design (IIIa, IIIb), and thus are expected to improve yield on devices meeting target operating wavelength specifications. 
Finally, we note that while aperiodically poled QPM designs (IIb and IIIb) are expected to yield devices with more ideal SH spectra (Fig.~\ref{main:fig3}b) and improve yield on devices meeting specifications, its advantages must be balanced against the significant metrology overhead (Supplementary Information) incurred compared to periodically poled designs (II/IIIa), which only require coarse mapping of the LN film thickness (e.g.~only one measured point near the middle of the waveguide). 

The preceding analysis makes the assumption that the measurement tools and strategies used are perfectly accurate and precise. 
However, in practice, both random and systematic errors can be present in measured data, thus providing inadequate simulation inputs. 
While random errors are hard to calibrate out and will result in a finite width of the measured SHG distribution, the systematic errors that result in displacement of the measured wavelength from the $737~\si{\nm}$ target can be calibrated out. 
This can be accomplished by poling and measuring a subset of calibration devices that are processed under identical etch and metrology conditions from the full sample to obtain a calibration curve that is used to adjust the fabricated grating periods on all remaining devices.

The calibration sample, fabricated on the same wafer, consists of a number of QPM waveguides with QPM grating period $\Lambda^\text{(sweep)}=2\pi/\qty(\abs{\Delta k^\text{(sim)}} + \Delta k_\text{sweep})$ which is intentionally swept, with $\Delta k_\text{sweep}$ taking on values from $-0.0785 \si{\micro\m}^{-1}$ to $+0.0785~\si{\micro\m}^{-1}$. 
Therefore, the expected SHG wavelength changes as a consequence of the varying sweep value, as shown by the orange markers in Fig.~\ref{main:fig3}d. 
In the case of perfect metrology, the measured SHG wavelength (Fig.~\ref{main:fig3}d, green markers) should be the same as the expected SHG wavelength. 
However, we find that the QPM grating spatial frequency, $K=2\pi/\Lambda$, for a device generating peak SH at $\lambda_\text{SH}$ is displaced from the simulated phase mismatch for that device by $\Delta k_\text{offset} \sim -0.645~\si{\micro\m}^{-1}$~(Fig.~\ref{main:fig3}d). 
This is a result of systematic errors that could be attributed to calibration errors in metrology or lithography tools, or inaccuracies in the index model used for the waveguide materials. 
To account for these errors, we correct the QPM periods of all devices on the same wafer as 
\begin{equation}
    \Lambda^\text{calibrated} = \frac{2\pi}{\abs{\Delta k^\text{(sim)}} + \Delta k_\text{offset}} 
    \label{eq:1}
\end{equation}
in order to ensure that the measured SHG spectrum peak coincides with the expected SHG wavelength.
Here, we assume that $\Delta k_\text{offset}$ is constant for all waveguides on the wafer. 

Using this calibration approach, we show that $73\%$ of devices out of $84$ measured (total of $280$ devices on Wafer~A fabricated without error and not used for calibration) support SHG of $(737 \pm 5)~\si{\nm}$ without additional tuning~(Fig.~\ref{main:fig3}, upper panel) despite implementing a simple periodic poling. 
Devices sampled from a separate $4''$ wafer run implementing aperiodic poling (Wafer~B) have a somewhat narrower distribution of operating wavelengths~(Fig.~\ref{main:fig3}e, lower panel). 
They are also blue-shifted by $\sim 4~\si{\nm}$ from the target wavelength, likely due to changes in metrology tool conditions during processing~(Supplementary Information). 
These devices are used in subsequent wavelength tuning and trimming experiments. 

\section*{Post-fabrication wavelength trimming and tuning}
Thermal tuning leveraging the thermo-optic effect in TFLN is the conventional approach to control the operating wavelength of QPM device \autocite{Hwang2023-kp,Liu2020-nz,Mishra2022-lb}. In our approach, we use a large heater to increase the temperature of Wafer B by $\Delta T_\text{max} = 120~\si{\K}$ above room temperature and measure the wavelength of SHG signal~(Fig.~\ref{main:fig4}a). 

This allows us to experimentally measure the thermo-optic tuning rate which is $\frac{\dd \lambda_\text{SH}}{\dd T} = 0.1~\si{\nm/\K}$.
Since increasing temperature can only red-shift the SHG wavelength, it is advantageous to start with blue-shifted QPM devices, like those found in our Wafer~B. 
Indeed, Figure~\ref{main:fig4}b shows that a device yield for Wafer B is $\sim 96\%$ at the maximum tested temperature. 
Here, we define device yield as the fraction of devices within tuning range.

In addition to cooling (not demonstrated here), we show that blue-shifting the operating wavelength of TFLN QPM devices can be achieved by depositing top cladding. 
We experimentally find that the operating wavelength can be blue-shifted by $\sim 50~\si{\nm}$ by depositing only $25~\si{\nm}$ of silica cladding. 
The blue-shifting effect saturates at silica cladding thicknesses above $200~\si{nm}$, since the optical mode is entirely contained within TFLN and silica, and its overlap with the air layer approaches zero (Fig.~\ref{main:fig4}c, lower panel). 
The trimming range of this approach ($>100~\si{\nm}$) is massive compared to thermo-optic tuning, and the amount trimmed can be very precisely controlled using atomic layer deposition. 
However, it must be noted that this cladding trimming approach will also modify waveguide dispersion, which can be inferred from the SHG spectra becoming narrower as more cladding is added (Fig.~\ref{main:fig4}c, upper panel). 
Thus for many dispersion-engineered QPM devices\autocite{Javid2021-ce, Mishra2022-lb, Wu2024-if, Ledezma2023-dy}, care must be taken when applying cladding trimming to ensure dispersion engineering goals are still met. 

\section*{Summary and outlook}
In conclusion, we develop an etch-before-pole process flow for QPM three-wave mixing devices optimized for scalability and wavelength-accurate operation within the limits of standard processing and metrology conditions. 
Wafer-scale fabrication of large arrays of QPM devices for SHG of $\lambda_0 = 737~\si{\nm}$ shows robustness to variations in etch depth and systematic errors in input parameters used in QPM grating design due to our calibration process. 
Where thermal tuning is possible, we demonstrate up to $96\%$ operating at the $737~\si{\nm}$ using an off-chip heater; future integration with efficient on-chip heaters \autocite{Liu2022-de} will allow individual devices to be tuned without affecting the remaining circuit. 
For applications where thermal tuning is not possible, such as interfacing with single-photon emitters at cryogenic temperatures \autocite{Wang2019-vm}, we also demonstrate cladding trimming to blue-shift operating wavelengths by $>100~\si{\nm}$. 
We note that by making further improvements in the accuracy of the ellipsometer and AFM, it may be possible to eliminate the systematic errors and may render calibration step and trimming unnecessary. 

Additionally, we also note that for the current poling electrode configuration used in our etch-before-pole approach, the expected maximum normalized SH conversion efficiency is expected to be reduced to $\sim 23\%$ compared to the conventional pole-before-etch approach because poling only occurs in the waveguide slab. 
Initial results that use a modified poling electrode geometry for the etch-before-pole approach to ensure that poling also occurs in the waveguide ridge have been demonstrated \autocite{Franken2024-ib}, and provide a promising path towards recovering ideal conversion efficiency. 

Improvements to wavelength-accurate QPM device yields in TFLN can accelerate the integration of QPM devices into more complex photonic integrated circuits or with other devices with relatively inflexible operating wavelength ranges, such as solid-state defect centers \autocite{Riedel2023-ks}. 
This can be especially helpful in the near- to medium-term while more precise fabrication techniques for the platform, such as atomic layer etching \autocite{Chen2023-li}, and process control best-practices are still under development. 
In particular, the SHG process chosen for the demonstrations in this work serves as a suitable proxy for identifying whether a device is a compatible frequency conversion module for $737~\si{\nm}$ emission from silicon-vacancy centers in diamond for most waveguide geometries in TFLN. Thus, this work has immediate applications in scalable quantum frequency conversion for realizing large-scale, low-loss, fiber-based quantum networks. 

\bigskip 

\section*{Acknowledgements}
This work was supported in part by 
AFOSR~FA9550-20-1-01015 (D.R., N.S.), 
A*STAR National Science Scholarship (S.L.), 
AWS Center for Quantum Networking’s research alliance with the Harvard Quantum Initiative (C.X., J.Y., A.S., L.M., D.R., M.L.), 
Ford Foundation Fellowship (D.R.), 
Harvard University Dean’s Competitive Fund from Promising Scholarship (M.L.), 
Harvard Quantum Initiative (N.A.), 
NASA~80NSSC22K0262 (J.Y., S.G., N.S.), 
NASA~80NSSC23PB442 (A.S.), 
NSF Engineering Research Center for Quantum Networks No.~EEC-1941583 (C.X., E.M., A.Z., N.S.), 
NSF~GRFP-1745303 (D.R.), 
NSF~OMA-2137723 (C.X., D.B.),  
NSF~OMA-2138068 (S.G., N.S.), and 
ONR~N00014-22-C-1041 (S.G., M.L.). 
Device fabrication was performed at the Center for Nanoscale Systems~(CNS), a member of the National Nanotechnology Coordinated Infrastructure Network~(NNCI), which is supported by the National Science Foundation under NSF Grant No.~1541959.

\section*{Author contributions}
C.X. developed the fabrication process, designed the devices and experiments. 
C.X. and B.D. optimized the ferroelectric poling process. 
A.S., S.L., and J.Y. fabricated devices and performed measurements with the assistance of N.A. 
C.X., S.L., and J.Y. analyzed the data. 
S.G. assisted with second harmonic imaging. 
D.R. and A.Z. assisted with the construction of the second harmonic imaging system. 
A.S. and L.M. developed the wafer scale process for etching lithium niobate. 
E.M. and D.B. assisted with developing film thickness and index measurement recipes. 
M.L. and N.S. supervised the project. 
C.X., S.L., J.Y., A.S., and M.L. wrote the manuscript in collaboration with the remaining co-authors. 
These authors contributed equally: C.X., S.L., J.Y., and A.S.

\section*{Competing interests}
M.L. is involved in developing lithium niobate technologies at HyperLight Corporation.
The remaining authors declare no competing interests.

\clearpage

\clearpage
\begin{figure}[h]
\caption{\justifying \bf Device geometry variation in a representative wafer run.}
\label{main:fig2}
\includegraphics[width=0.5\textwidth]{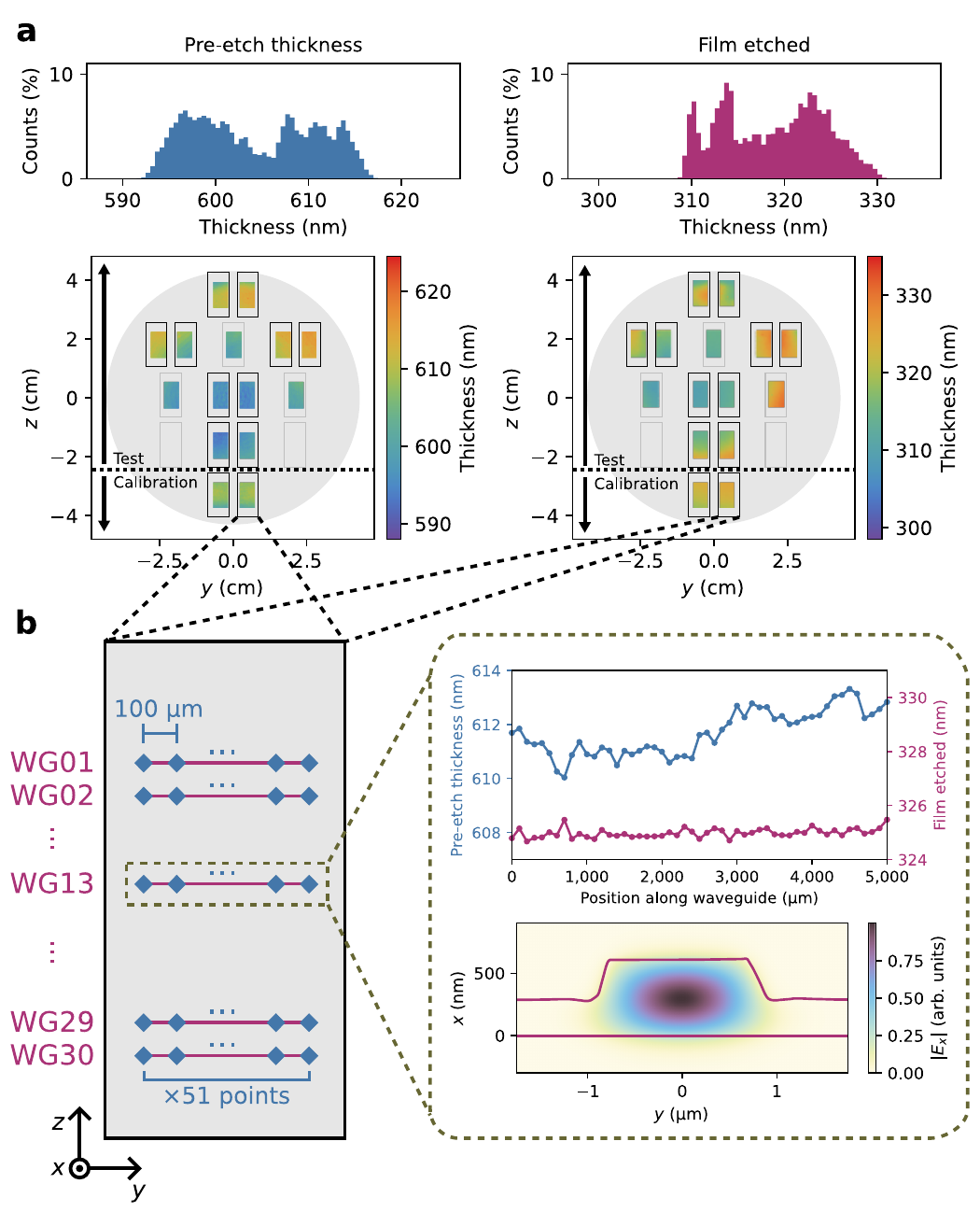}
\caption*{\justifying
    \textbf{a},~Non-contact optical measurements of LN film thickness before etching (left) and the amount etched (right) from a representative wafer run.
    The amount etched is calculated from the film thickness difference between pre- post-etch wafer maps (see Supplementary Information for post-etch data) and shows global variation on the order of $\sim 20~\si{\nm}$.
    Black outlines indicate locations of dies containing QPM devices relevant to this work; locations of dies corresponding to other work are outlined in light gray for completeness.
    Devices below the horizontal dotted line are used for systematic error calibration, and thus are poled and measured prior to the remaining test devices on the wafer.
    \textbf{b},~Not-to-scale diagram of film thickness measurement locations on a QPM device die. Blue markers indicate the location of scan points relative to QPM devices, represented by purple lines (left).
    Local pre-etch film thickness and etch depth variation along a single waveguide (top right).
    Simulated fundamental transverse electric (TE00) mode profile at the target second harmonic (SH) wavelength, $\lambda_\text{target} = 737~\si{\nm}$; waveguide profile from the nearest AFM measurement site is used in the simulation (lower right).
    Scan pattern shown is for a wafer implementing aperiodic QPM designs; a less dense scan pattern is used for a separate wafer run implementing periodic QPM to reduce metrology overhead (Extended Data Fig.~\ref{ext:fig1}). 
}
\end{figure}

\clearpage
\begin{figure}[h]
  \caption{\justifying \bf Estimated and measured device operating wavelength distributions.}
  \label{main:fig3}
  \includegraphics[width=\textwidth]{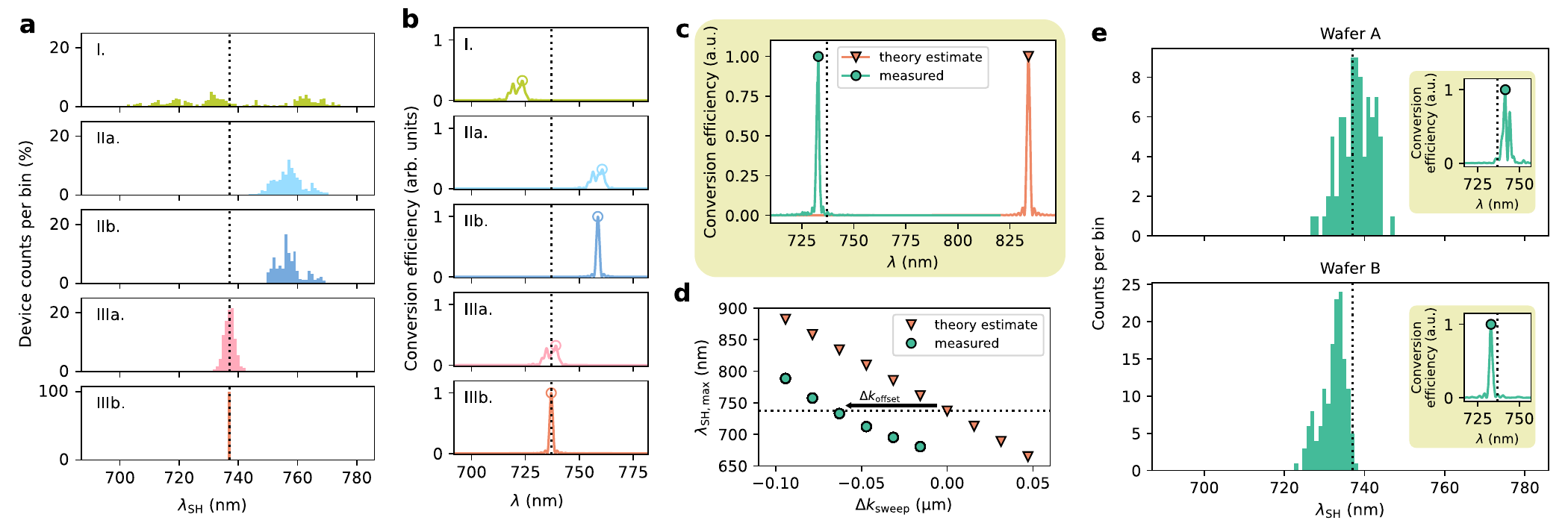}
  \caption*{\justifying
        \textbf{a}, Simulated distribution of $\lambda_\text{SH}$ for the $N = 360$ QPM devices in the wafer pictured in Fig.~{\ref{main:fig2}a} for five different types of QPM grating. 
        The QPM grating design methods considered are: 
            (I)~periodic poling based on ideal waveguide cross-section geometry ($t = 600~\si{\nm}$,  $h = 300~\si{\nm}$, $w = 1.5~\si{\micro\m}$, $\theta = \ang{60}$); 
            (IIa)~periodic poling using a single-point pre-etch film thickness measurement;
            (IIb)~locally varying period calculated using data at each film thickness measurement location; 
            (IIIa)~periodic poling based on the measured waveguide geometry at a single point along the waveguide, and 
            (IIIb)~locally varying period calculated using measured waveguide geometry at each film thickness measurement location. 
        Methods I and II can be used in the conventional pole-before-etch process, whereas Method III is only possible for the etch-before-pole process. 
        Dotted lines indicate the position of $\lambda_0$.
        \textbf{b},~Example SH spectra for each QPM design method considered in (\textbf{a}). 
        IIa (IIIa) has a wider spread relative to IIb (IIIb) due to a more displaced $\lambda_\text{SH}$ (markers). 
        \textbf{c},~Comparison of estimated and measured SH spectra for a calibration device with QPM grating designed using Method~IIIb.
        Simulated phase mismatches are given a swept offset, $\Delta k_\text{sweep}$, in the calibration device (Eq.~\ref{eq:1}) to deliberately produce a QPM design where $\lambda_\text{SH}$  is red-shifted relative to $\lambda_0$ (orange curve).
        Measured data (green curve) reproduces the ideal sinc-squared SH spectrum well but is blue-shifted relative to the theoretical estimate. 
        \textbf{d},~Estimated and measured $\lambda_\text{SH}$  for a set of calibration devices. Estimation lacks curvature present in the measured data due to first-order estimation (Methods). 
        Dotted line indicates the position of ${\lambda_0}$, while the arrow indicates $\Delta k_\text{offset}$ used for calibration of remaining devices on the same wafer. 
        \textbf{e},~Distributions of measured $\lambda_\text{SH}$ for test devices on two wafer runs relative to $\lambda_0$ (dotted line) after systematic error calibration is applied. $n = 84 $ ($n = 144$) devices are sampled for periodically poled Wafer~A (aperiodically poled Wafer~B) out of a population of $N=280$ ($N=240$) non-calibration devices without known non-poling processing defects~(Methods). 
        The total number of non-calibration devices fabricated in Wafer A (B) is $600$ ($300$). 
    }
\end{figure}

\clearpage
\begin{figure}[h]
    \caption{\justifying\bf Device yields vis-\`a-vis tunability of target three-wave mixing process.}
    \label{main:fig4}
    \includegraphics[width=0.5\textwidth]{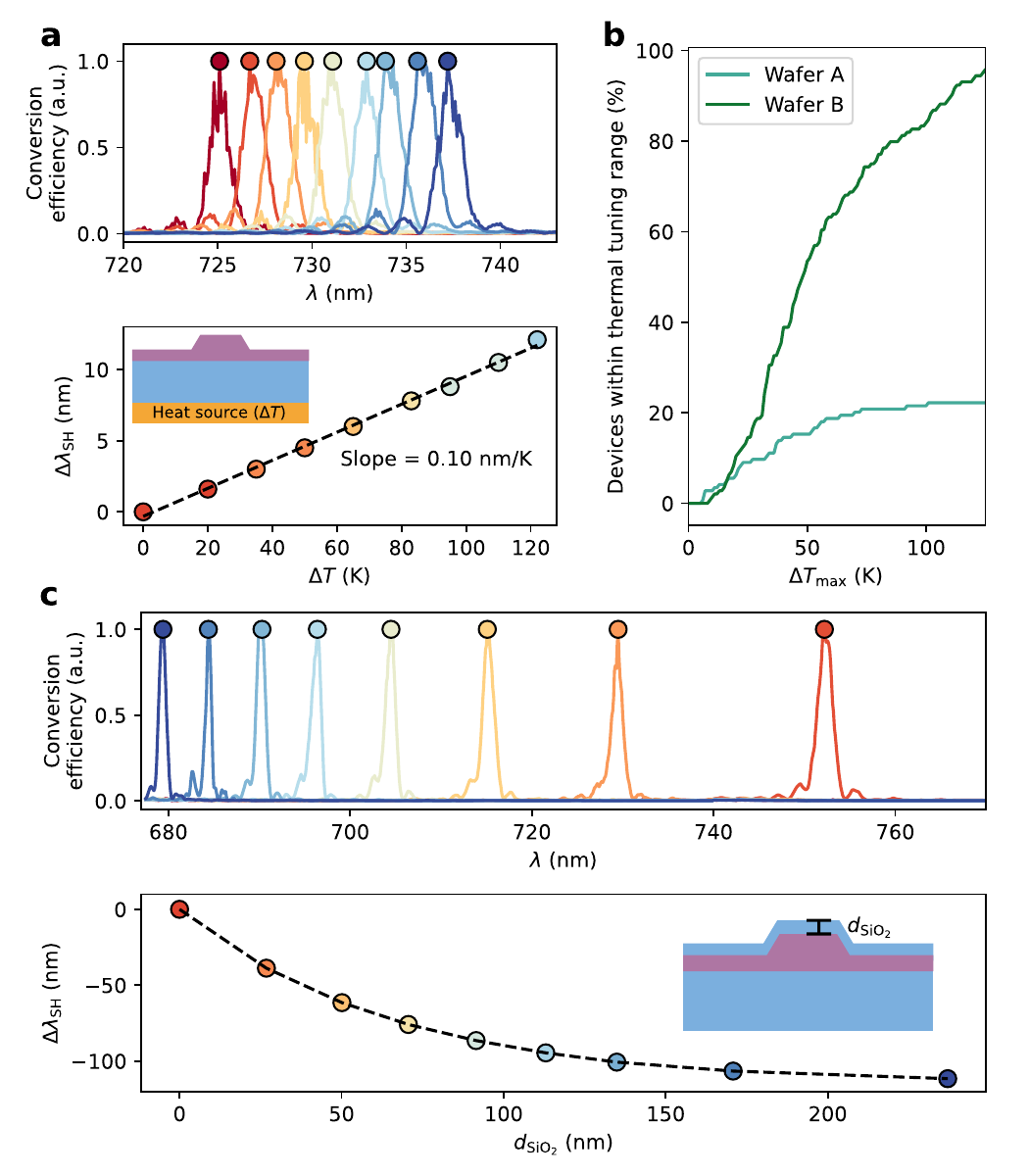}
    \caption*{\justifying
        \textbf{a},~Thermo-optic tuning up to $\Delta T_\text{max} = 120\si{\degreeCelsius}$ above room temperature. 
        SH spectra maintain approximately the same shape over the tuning range (upper panel). 
        Red-shifting of up to $12.1~\si{\nm}$ is demonstrated (lower panel) using an off-chip ceramic heater (inset). 
        \textbf{b},~Yield on devices with operating wavelengths within thermo-optic tuning range of $\lambda_0$ as a function of the maximum achievable temperature for measured distributions in Fig.~\ref{main:fig3}d. 
        \textbf{c},~Tuning of device operating wavelength via deposition of thin layers of cladding oxide, $d_{\rm SiO_2} $(lower panel inset). 
        Experimental data showing $>100~\si{\nm}$ blue-shift of the operating wavelength observed, with the rate of blue-shift decreasing as more silica cladding is deposited (lower panel). 
        At large cladding thicknesses, widths of the SH spectra decrease as operating wavelengths are blue-shifted, indicating increasing group velocity mismatch between the fundamental and second harmonic of the quasi-phase matched process (upper panel); the impact of this is minor, however, for the small amounts of wavelength trimming ($<50~\si{\nm}$) necessary in most cases. 
    }
\end{figure}

%\\\\
\clearpage \printbibliography

\clearpage

\section*{\large Methods}

\subsection*{Device fabrication}
Devices are fabricated on commercially available $4''$ thin-film lithium niobate on insulator wafers (NanoLN). 
The nominal material stack consists of a $600~\si{\nm}$ device layer ($x$-cut 5\% MgO-doped lithium niobate) on $2~\si{\micro\m}$ of buried oxide on a silicon handle. 
Prior to processing, variable angle spectroscopic ellipsometry is used to measure the refractive index of the material as a function of wavelength. 

Alignment markers are first patterned on the wafer via photolithography and metal lift-off. 
An aligned ellipsometry step is then carried out to produce a map of device layer film thickness. 
For periodic poling, film thickness is sampled with a step size of $1~\si{\mm}$ in between pairs of adjacent QPM waveguides (Extended Data Fig.~\ref{ext:fig1}) to minimize metrology overhead. 
For aperiodic poling, film thickness is sampled with a step size of $100~\si{\micro\m}$ along each QPM waveguide (Fig.~\ref{main:fig2}b). 

Following pre-etch metrology, a hard mask of $600~\si{\nm}$ of silicon dioxide is deposited on the wafer via plasma-enhanced chemical vapor deposition (PECVD). 
Electron beam lithography is then used to pattern the waveguide layer using a negative tone resist (ma-N~2400). 
The pattern is then transferred to the hard mask and the device layer via dry etching. 
The nominal etch depth of $300~\si{\nm}$ is targeted by calibrating the etch time in situ. 

Post-etch ellipsometry is carried out to map the amount of remaining slab in the device layer at the same points sampled during the pre-etch measurement. 
Remaining waveguide geometry parameters, such as etch angle and measured top width, are extracted by AFM on a witness feature on each QPM device die (each comprising $60$ QPM devices). 
A high-aspect ratio probe (Nanosensors~AR5-NCLR) is used to minimize the impact of distortions introduced by the probe tip profile. 
The measured waveguide geometry parameters are used in FDE simulations to calculate local variations in the expected phase mismatch of the target $\chi^{(2)}$ frequency conversion process. 

Coplanar finger electrodes (Fig. 1b, inset) for ferroelectric poling are patterned via electron beam lithography using standard positive resist (PMMA~C4) and metal lift-off ($130~\si{\nm}$ gold on $25~\si{\nm}$ chromium). 
Following metal lift-off, a cap layer of photoresist (Microposit S1800 series) is spun on to prevent dielectric breakdown through air during poling. 
Contact pad windows are lithographically defined to allow probing. 
High-voltage pulses are applied to the finger electrodes to create the QPM grating for the target $\chi^{(2)}$ frequency conversion process. 
Second harmonic microscopy is used to validate the poling quality. Electrodes are removed via wet etch prior to measurement. 

In each wafer run, the small wafer segment containing calibration devices is cleaved or diced from the full wafer after waveguides are etched so that they can be poled and measured separately from the test devices. 
All devices in the remaining segment are processed simultaneously after the calibration parameter is extracted from measurements of the calibration devices. 
For cladding trimming, the same PECVD silica was deposited in $\sim 25~\si{\nm}$ increments. 

\subsection*{Waveguide cross-section geometry metrology}
Pre- and post-etch film thickness measurements are carried out using a variable-angle spectroscopic ellipsometer (SE). 
Film thickness map measurements were carried at a fixed angle of $\ang{55}$. 
Focusing probes are used to reduce the spot size to $<100~\si{\micro\m}$ to resolve fine local film thickness variations for aperiodic poling. 
Film thickness data is extracted from raw ellipsometric data by fitting to in-house index models for both the device and buried oxide layers. 
The in-house index models are developed by fitting multi-angle SE measurements of an as-received wafer to Sellmeier index models in the $500~\si{\nm}$ to $2,500~\si{\nm}$ wavelength range. 

We note that both ellipsometry and AFM provide measurements of etch depth, $d$. 
Under ideal tool conditions, the etch depth extracted from these two measurement approaches should give the same results.  
However, we find that this is generally not the case, which indicates the existence of calibration errors. 
We find that data obtained from ellipsometry results in better predictions of the SHG wavelength. 
For this reason, we also use ellipsometry measured etch depth to calibrate the AFM obtained one, which also calibrates that sidewall measurement obtained by AFM (waveguide width AFM measurements are not affected).

\subsection*{Simulation of phase-mismatch at the target three-wave mixing process}
After calibration, ellipsometry and AFM data are combined to produce a complete waveguide cross-section profile (Fig.~\ref{main:fig2}b, lower right) for FDE simulations~(Lumerical MODE). 
FDE simulations are performed at the second harmonic wavelength $\lambda_\text{SH} = \lambda_0$ and fundamental harmonic wavelength $\lambda_\text{FH} = 2\lambda_0$ to find the phase mismatch, 
$\Delta k^\text{(sim)} = k_\text{SH} - 2k_\text{FH}$, where $k_\text{SH(FH)}$ is the propagation constant of the TE00 mode at the SH (FH) wavelength. 

One AFM measurement is carried for per set of $60$ QPM devices due to AFM being significantly more time intensive. 
For each set of devices that share the same AFM waveguide profile, only pre- and post-etch LN thickness from the ellipsometry from point-to-point. 
For aperiodic poling, $3060$ points are measured for each set of devices. An exhaustive simulation at the requisite FDE mesh size ($5~\si{\nm} \times 5~\si{\nm}$) is computationally intensive. 
To mitigate this, the $2$-dimensional parameter space of post-etch LN thickness and the etch depth for devices using the same AFM profile is binned into a 2-dimensional histogram with $0.5~\si{\nm} \times 0.5~\si{\nm}$ bins. 
The coordinates of the vertices of each non-empty bin are recorded and simulated (Extended Data Fig.~\ref{ext:fig2}a), reducing the number of simulations for similar geometries. 
Since $\Delta k^\text{(sim)}$ is smooth and continuous, bilinear interpolation based on simulation results at the vertex coordinates (Extended Data Fig.~\ref{ext:fig2}b) can be used to calculate $\Delta k^\text{(sim)}$ at each measured point (Extended Data Fig.~\ref{ext:fig2}c).  

\subsection*{Theoretical estimates of SH spectra}
Theoretical estimates of the SH spectra, such as those in Fig. 3a, are numerically calculated using FDE-simulated propagation constants, $k_\text{eff} = k_\text{eff}(\lambda,y)$ of the TE00 mode at wavelength $\lambda$ using the measured waveguide cross-section geometry at position $y$ along a QPM device the method described in Ref.~\cite{Chen2024-yt}. 
Fluctuations in $k_\text{eff}$   below the the film thickness map resolution of $\Delta y = 100~\si{\micro\m}$ are ignored. 
To ensure $\lambda_\text{SH}$ can be correctly identified for a wide range of geometry parameters, a wide simulation range of $\lambda_\text{SH} = 600~\si{\nm}$ to $850~\si{\nm}$ was chosen. 
To reduce simulation complexity for the large number of waveguide geometries in each wafer run, $k_\text{eff}$ is only simulated for two wavelengths: $\lambda_\text{SH,0} = \lambda_0$ and $\lambda_\text{FH, 0} = 2\lambda_0$. 
For SH (FH) wavelengths other than $\lambda_\text{SH(FH),0}$, a first-order estimate of $k_\text{eff}$ is used:
\begin{equation*}
    k_\text{eff}\qty(\lambda_\text{SH(FH)}, y) \approx k_\text{eff} \qty(\lambda_\text{SH(FH), 0}, y) - \frac{2\pi n_g\qty(\lambda_\text{SH(FH),0}, y)}{{\lambda_\text{SH(FH),0}}^2} \Delta \lambda, 
\end{equation*}
where $\Delta\lambda = \lambda_\text{SH(FH)} - \lambda_\text{SH(FH), 0}$, and the group index, $n_g\qty(\lambda_\text{SH(FH)},0, y)$, is also extracted from the FDE simulation at $\lambda_\text{SH(FH), 0}$. 
At large $\Delta\lambda$, higher-order dispersion terms must be taken into account for a more accurate calculation of $k_\text{eff} (\lambda_\text{SH(FH)},y)$, and consequently $\lambda_\text{SH}$. 
This is evident in Fig.~\ref{main:fig3}d, where the theoretical estimate of $\lambda_\text{SH}$ does not reflect the curvature present in the measured data. 
Notwithstanding, the first-order estimate is sufficient for identifying qualitative differences in the $\lambda_\text{SH}$ for different methods of designing the QPM grating (Fig.~\ref{main:fig3}b). 

\subsection*{Measurement of SH spectra}
The SH spectrum of a QPM device is measured by measuring the SH power output as a function of input FH wavelength. 
Two FH sources are combined for continuous wavelength coverage from the telecommunications E-band (Santec TSL-550; $1355~\si{\nm}$ to $1485~\si{\nm}$) to L-band (Santec TSL-710; $1480~\si{\nm}$ to $1640~\si{\nm}$). 
Light is coupled into and collected from the device under test (DUT) using lensed fibers (OZ Optics). 
A fiber polarization controller at the output of each FH laser is used to ensure TE input into the device under test (DUT) in order to measure the target type-0 SHG process. 
The FH wavelength is stepped with a resolution of 0.2 nm and the SH power is recorded at each FH wavelength step using a fiber-coupled visible power meter (Thorlabs S150C). 
A reference curve of the FH power at the DUT output as a function wavelength is also acquired (Thorlabs S155C) in a separate sweep following the SH spectrum measurement. 

All measurements are performed at room temperature with the exception of the thermal tuning experiment. 
The thermal tuning experiment is performed by heating the device die using an off-chip $20~\si{\W}$ metal ceramic heater; device temperature is measured using a $100~\si{\ohm}$ platinum resistance temperature detector.

\subsection*{Device sampling}
The distribution of operating wavelengths of devices in each wafer run is estimated by measuring the SH spectra of a subset of the population of test devices without known non-poling processing defects.
An accounting of the total number of devices fabricated in each wafer run, and known processing defects is provided below: 

\subsubsection*{Wafer~A}
A total of 660 devices relevant to this work are fabricated in this wafer run, with $600$ ($60$) devices designated as test (calibration) devices. 
Of the test devices, 40 devices with anomalous top widths (Extended Data Fig.~\ref{ext:fig3}a) due to stitching errors in the electron beam lithography step are excluded from consideration. 
A further $280$ devices’ waveguide profiles are measured with a different AFM tip than the calibration devices’ due to damage to the original tip, and are thus excluded since the assumption of consistent systematic errors does not hold. 
The population of test devices without known non-poling processing defects in Wafer A is $N=280$; the location of excluded devices is indicated in Extended Data Fig.~\ref{ext:fig3}b. 
During measurement, a total of $n=84$ ($35\%$) devices are sampled to estimate the distribution of $\lambda_\text{SH}$ shown in Fig.~\ref{ext:fig3}e (upper panel). 
A subset of devices measured with a different AFM tip were tested to verify that the systematic error from waveguide profiling changes with AFM tip used (Extended Data Fig.~\ref{ext:fig3}c).

\subsubsection*{Wafer~B}
A total of $360$ devices relevant to this work are fabricated in this wafer run, with $300$ ($60$) devices designated as text (calibration) devices. 
Of the test devices, an error in the extraction of the post-etch film thickness data during QPM design resulted in anomalous $\lambda_\text{SH}$ for $60$ devices. 
The error in the post-etch film thickness is identified by comparing the fine wafer map data against a reference coarse wafer map (Extended Data Fig.~\ref{ext:fig3}d). 
The correct post-etch film thickness is subsequently extracted during post-mortem analysis  (Extended Data Fig.~\ref{ext:fig3}d, right panel), and the devices designed with erroneous film thickness data are excluded from the population of the valid test devices. 
The population of test devices without known non-poling processing defects in Wafer A is $N=240$. During measurement, a total of $n=144$ ($60\%$) devices are sampled to estimate the distribution of $\lambda_\text{SH}$. 

\renewcommand\figurename{Extended Data Figure}
\setcounter{figure}{0}    

\clearpage
\begin{figure}[htbp]
    \caption{\justifying \bf
        Film thickness map for Wafer~A implementing periodic poling.
    } \label{ext:fig1}
    \includegraphics[width=\textwidth]{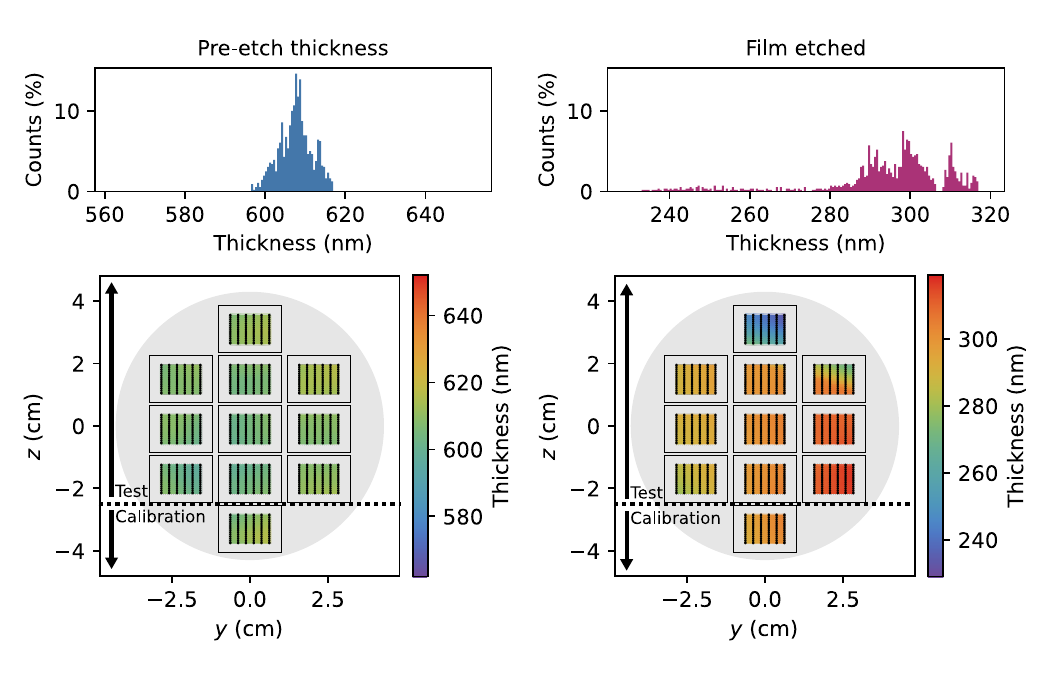}
    \caption*{\justifying
        Pre-etch film thickness and amount of film etched for wafer run implementing periodic QPM grating designs (Wafer~A). 
        The number of measured points---and therefore metrology overhead---is reduced by an order of magnitude by sampling along each device with a step size of $1~\si{\mm}$ rather than the $100~\si{\micro\m}$ step size used for Wafer~B, which implements aperiodic QPM grating designs (Fig.~\ref{main:fig2}b).
    }
\end{figure}

\clearpage
\begin{figure}[htbp]
    \caption{\justifying \bf
        Simulation of phase mismatch for target SHG process.
    } \label{ext:fig2}
    \includegraphics[width=\textwidth]{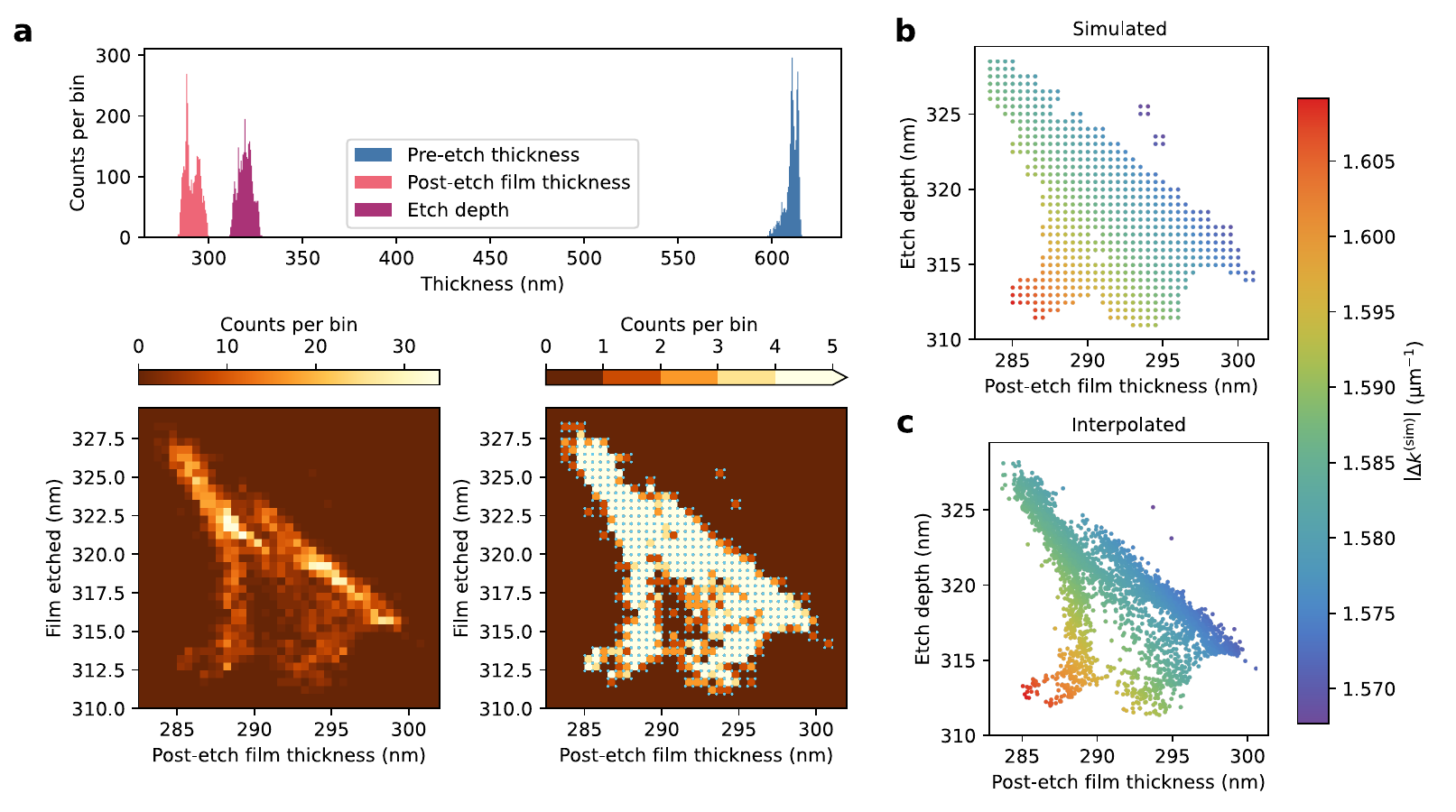}
    \caption*{\justifying
        \textbf{a},~Distributions of pre- and post-etch LN thickness and etch depth for a set of $60$ QPM devices using the same AFM cross-sectional data for simulation (upper panel). 
        A $2$-D distribution of post-etch thickness and etch depth (lower left). 
        Phase mismatch for the target process, $\Delta k^\text{(sim)}$, is simulated for the coordinates at the corners of each occupied bin (lower right, markers). 
        \textbf{b},~$\Delta k^\text{(sim)}$ at selected coordinates are smooth and continuous and therefore can be interpolated. 
        \textbf{c},~Bilinear interpolation of simulated points to the each measured waveguide geometry is used to extract the $\Delta k^\text{(sim)}$ values used to design QPM gratings. Each point in the scatter plot represents a measured waveguide geometry, which is defined by its slab thickness (post-etch thickness) and rib height (etch depth). 
    }
\end{figure}

\clearpage
\begin{figure}[htbp]
    \caption{\justifying \bf
        Non-poling processing defects.
    } \label{ext:fig3}
    \includegraphics[width=\textwidth]{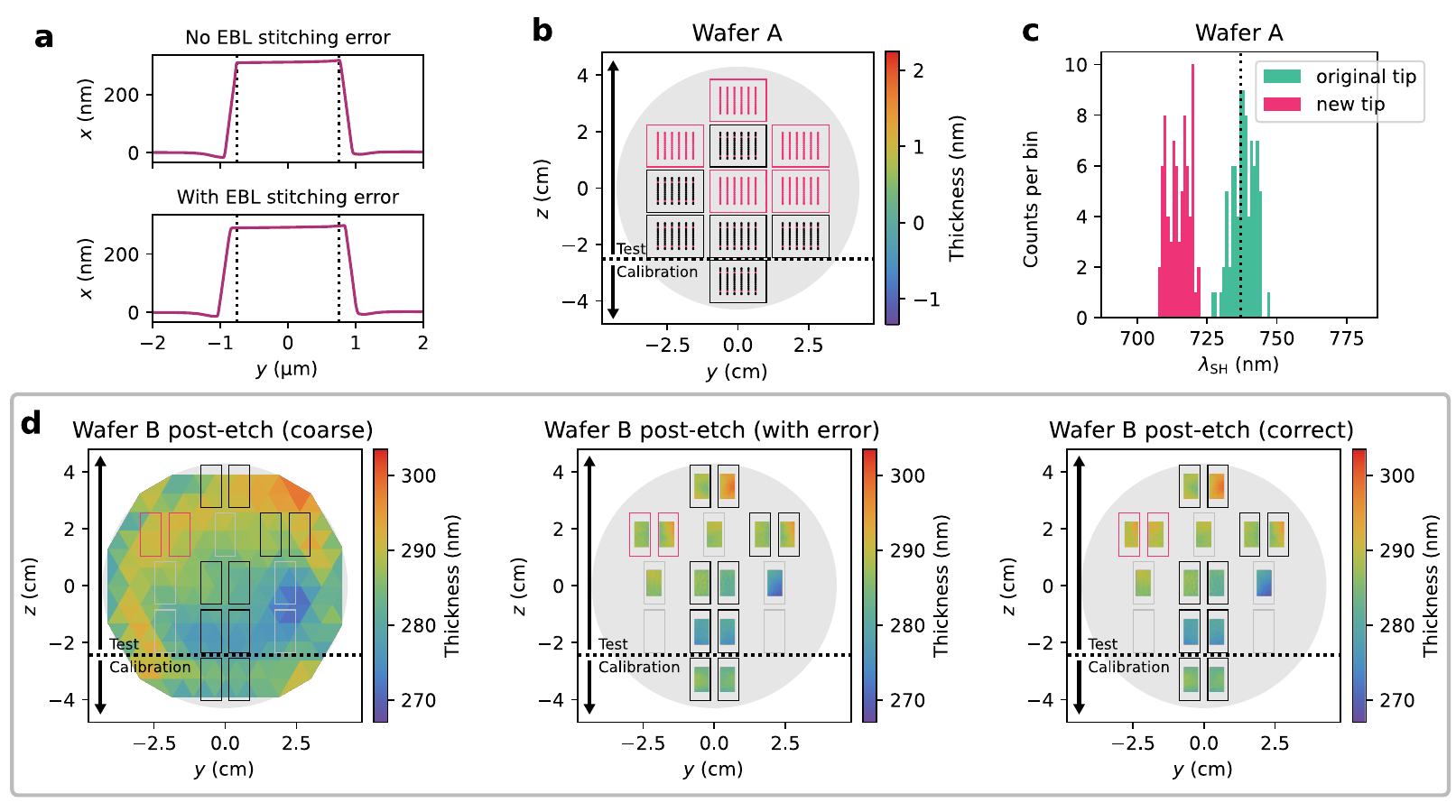}
    \caption*{\justifying
        \textbf{a},~Comparison of AFM profile of waveguides on Wafer A with and without EBL stitching error. The waveguide top width is noticeably dilated relative to the nominal top width (dotted lines) for waveguides affected by stitching error (lower panel), whereas unaffected waveguides have close to nominal top width (upper panel). 
        \textbf{b},~Wafer A devices in locations highlighted in magenta feature non-poling processing defects and are excluded from the aggregate shown in Fig.~\ref{main:fig3}e (upper panel). 
        All devices located within magenta boxes are excluded because a new AFM tip had to be used for waveguide profiling instead of the original tip used to profile the calibration devices, and therefore feature different systematic errors compared to the calibration devices. 
        \textbf{c},~$n=75$ ($27\%$) devices with waveguide profiles taken using the new AFM tip are measured. We verify that the distribution $\lambda_\text{SH}$ for these devices are indeed shifted relative to the correctly calibrated devices (original tip), reflecting a different systematic error introduced by using the new tip.
    }
\end{figure}

\clearpage
\begin{figure}[htbp]
    \caption{\justifying \bf
        Impact of reduced overlap between poled region and optical modes on conversion efficiency.
    } \label{ext:fig4}
    \includegraphics[width=0.5\textwidth]{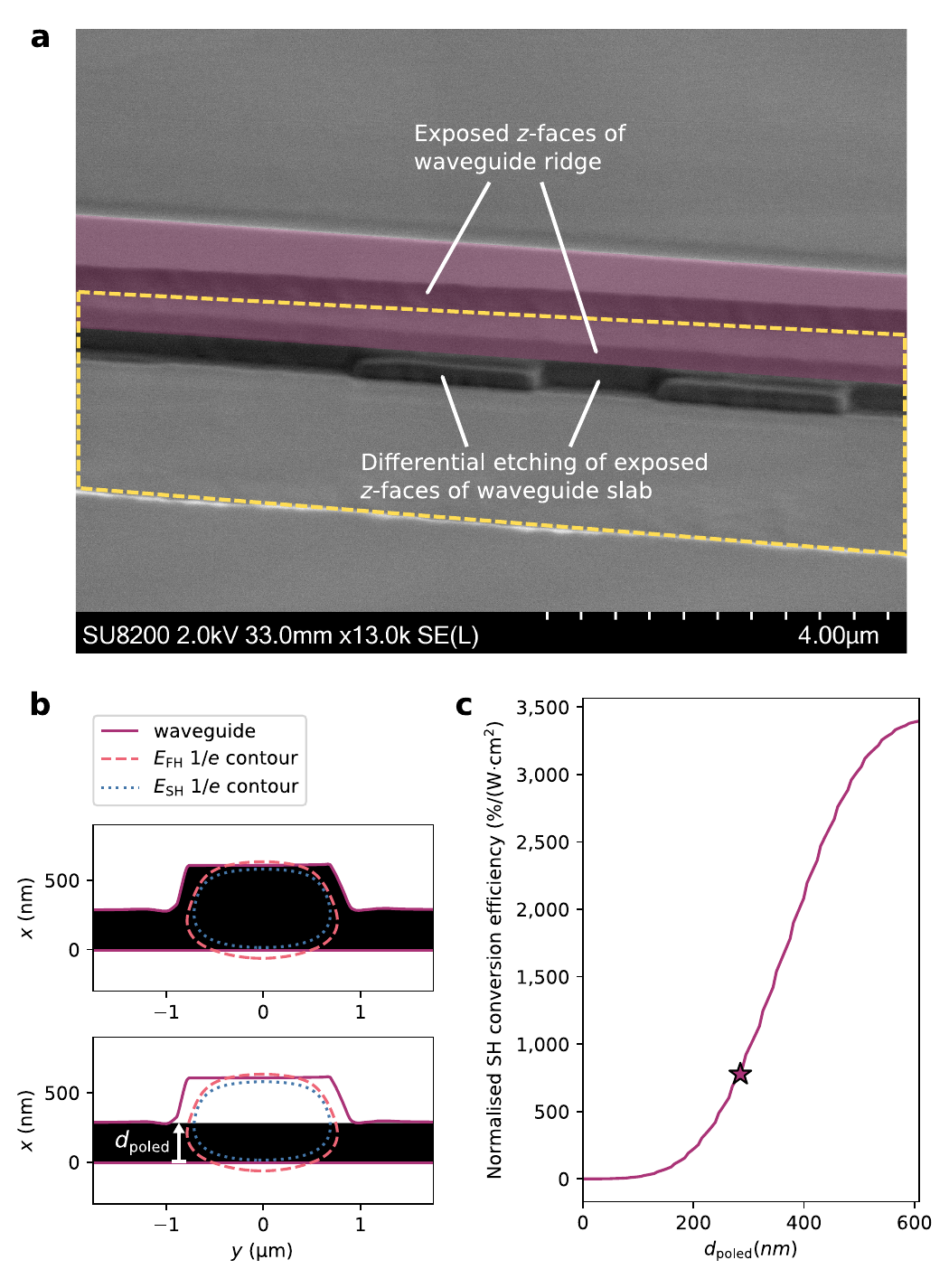}
    \caption*{\justifying
    \textbf{a},~False-color scanning electron micrograph of an etch-before-pole waveguide after dry-etching to expose LN $z$-faces and wet etching using $49\%$ hydrofluoric acid (HF) to expose structure of inverted domains. 
    The region that is dry-etched to expose LN $z$-faces is bounded by dashed yellow lines. 
    The position of the waveguide ridge is highlighted in purple. 
    Only the region immediately below the waveguide ridge, corresponding to the waveguide slab, shows corrugation indicative of differential etching of $+z$- and $-z$-faces in HF. 
    This shows that domain inversion in the etch-before-pole process only propagates through the waveguide slab. 
    \textbf{b},~Measured waveguide profile along with the simulated $1/e$ contours of the fundamental TE SH (FH) electric fields, $E_\text{SH(FH)}$, for a fully poled waveguide (upper panel) and a partially poled waveguide (lower panel) with the poled region shaded in black.
    $d_\text{poled}$ indicates the thickness of poled LN measured from the bottom of the LN film. 
    \textbf{c},~Estimated normalised conversion efficiency for the target SHG process as a function of $d_\text{poled}$. 
    Marker indicates the normalised conversion efficiency for a configuration in which only slab of the waveguide is poled for the waveguide geometry shown in~(\textbf{b}). 
    }
\end{figure}
\end{document}